\documentclass[pre,floats,floatfix,twocolumn,aps,superscriptaddress]{revtex4}

\usepackage[pdftex]{graphicx}
\usepackage[pdftex]{color}
\usepackage{dcolumn}
\usepackage[english]{babel}
\usepackage{lmodern}
\usepackage[latin2]{inputenc}
\usepackage[T1]{fontenc}
\usepackage{textcomp}
\usepackage{ae} %
\usepackage{amssymb}
\usepackage{mathrsfs}
\usepackage{verbatim}
\usepackage{latexsym}
\usepackage{babelbib}
\usepackage{bpchem}
\usepackage[version=3]{mhchem}
\usepackage{url}
\usepackage[tight,nice]{units}
\usepackage{booktabs}

\usepackage{fancyhdr}
\usepackage[justification=RaggedRight,singlelinecheck=false,small,bf]{caption} 

\usepackage{syntonly}

\usepackage[pdftex,bookmarks,pdftitle={Draft},pdfauthor={Anze Bozic},pdfsubject={None},pdfdisplaydoctitle,pdfstartview=FitH,unicode=true]{hyperref}

\begin{document}


\title{How simple can a model of an empty viral capsid be? Charge distributions in viral capsids}

\author{An\v{z}e Lo\v{s}dorfer Bo\v{z}i\v{c}\footnotemark[1], Antonio \v{S}iber\footnotemark[2], and Rudolf Podgornik\footnotemark[1]\footnotemark[3]}
\affiliation{\footnotemark[1]Department of Theoretical Physics, Jo\v{z}ef Stefan Institute, SI-1000 Ljubljana, Slovenia\\\footnotemark[2]Institute of Physics, Bijeni\v{c}ka cesta 46, P.O. Box 304, 10001 Zagreb, Croatia\\\footnotemark[3]Department of Physics, Faculty of Mathematics and Physics, University of Ljubljana, SI-1000 Ljubljana, Slovenia}

\begin{abstract}
We investigate and quantify salient features of the charge distributions on viral capsids. Our analysis combines the experimentally determined capsid geometry with simple models for ionization of amino acids, thus yielding the detailed description of spatial distribution for positive and negative charge across the capsid wall. The obtained data is processed in order to extract the mean radii of distributions, surface charge densities and dipole moment densities. The results are evaluated and examined in light of previously proposed models of capsid charge distributions, which are shown to have to some extent limited value when applied to real viruses.
\end{abstract}

\date{\today}
\maketitle

\section{Introduction}

Starting from the early structural studies of tobacco mosaic virus gels, Bernal and Fankuchen~\cite{Bernal} already invoked electrostatic interactions that are ``probably due to the ionic atmospheres surrounding [viruses]'' to explain their  behavior in ionic solutions. Virus architecture, cell attachment, penetration, progeny assembly and egress should be dependent on long-range colloidal interactions between and within viruses and various other structural components of the cell \cite{Iwasaki}. Though the importance of electrostatic interactions in the context of viruses is well recognized~(see the review by \v{S}iber et al.~\cite{Siber2012} and references therein) and electrostatic models on various levels of sophistication abound \cite{Siber2007,Prinsen2010,ALB2011,Marzec1993,Zandi2006,Kegel2004,Kegel2006,Belyi2006,Ting2011}, preciously little systematic effort \cite{Karlin1988,Michen2010} has been directed towards detailed quantification of the charge distributions on and within viral capsids. Models of electrostatic interactions in the context of viruses as well as virus-like nanoparticles \cite{Nicole,nanospheres} only make sense if they are derived from detailed observed charge distributions on the epitopal and hypotopal surfaces\footnote{Outer and inner surfaces, respectively.} of the capsid, as well as charge buried inside the capsomeres. Therefore, to evaluate previous modeling attempts, to propose better models, and to find out whether there is a prototypical charge distribution of a virus capsid, we embark on a detailed study of charge distribution on empty viral capsids. 

Our focus will not reside upon the distribution of charged amino acids along the 1D primary sequences of capsomeres \cite{Karlin1988} but exclusively on the 3D geometry of the charge distribution on the capsid.  While the details of the large-scale nature of the electronic structure of proteins that would allow the assessment of partial charge distribution buried inside the protein core are presently unavailable \cite{WYChing2011,Pichierri}, the charges of the amino acids residing on the surface of the capsomers in contact with the aqueous solvent at physiological \textsl{p}H are known and readily available \cite{Ptitsyn2002}. We will use the charge distribution on the epitopal and hypotopal capsid surfaces of a large number of viruses in order to analyze and model its statistical signature among the various virus types. 

In order to describe any charge distribution one first needs to identify the spatial region in which such a distribution resides and then quantify its geometry via a set of lowest multipolar moments \cite{Jackson1999}. With this goal in mind we will examine a number of available X-ray scattering and cryo-electron microscopy structural data on capsids of various viruses in order to extract a small set of parameters that would characterize simple models of charge distribution pertaining to these capsids. This minimal set of parameters includes the average size and thickness of the capsid, the surface charge density, and surface dipole density magnitude of the charge distribution.

The structure of the paper is as follows: We first explain how we construct two simple capsid models from the experimental data and obtain the parameters pertaining to them. We then briefly analyze the geometrical properties of the two models before proceeding to the monopolar and dipolar charge distributions on the capsids. We focus on different surface charge distributions pertaining to both models, and the effect of charge on the disordered protein N-tails. Lastly, we consider the surface dipole density in capsids, and conclude with the discussion of our results.

\section{From Structures to Model(s)}
\label{sec:s2m}

\begin{figure}[!htp]
\begin{center}
\includegraphics[width=\columnwidth]{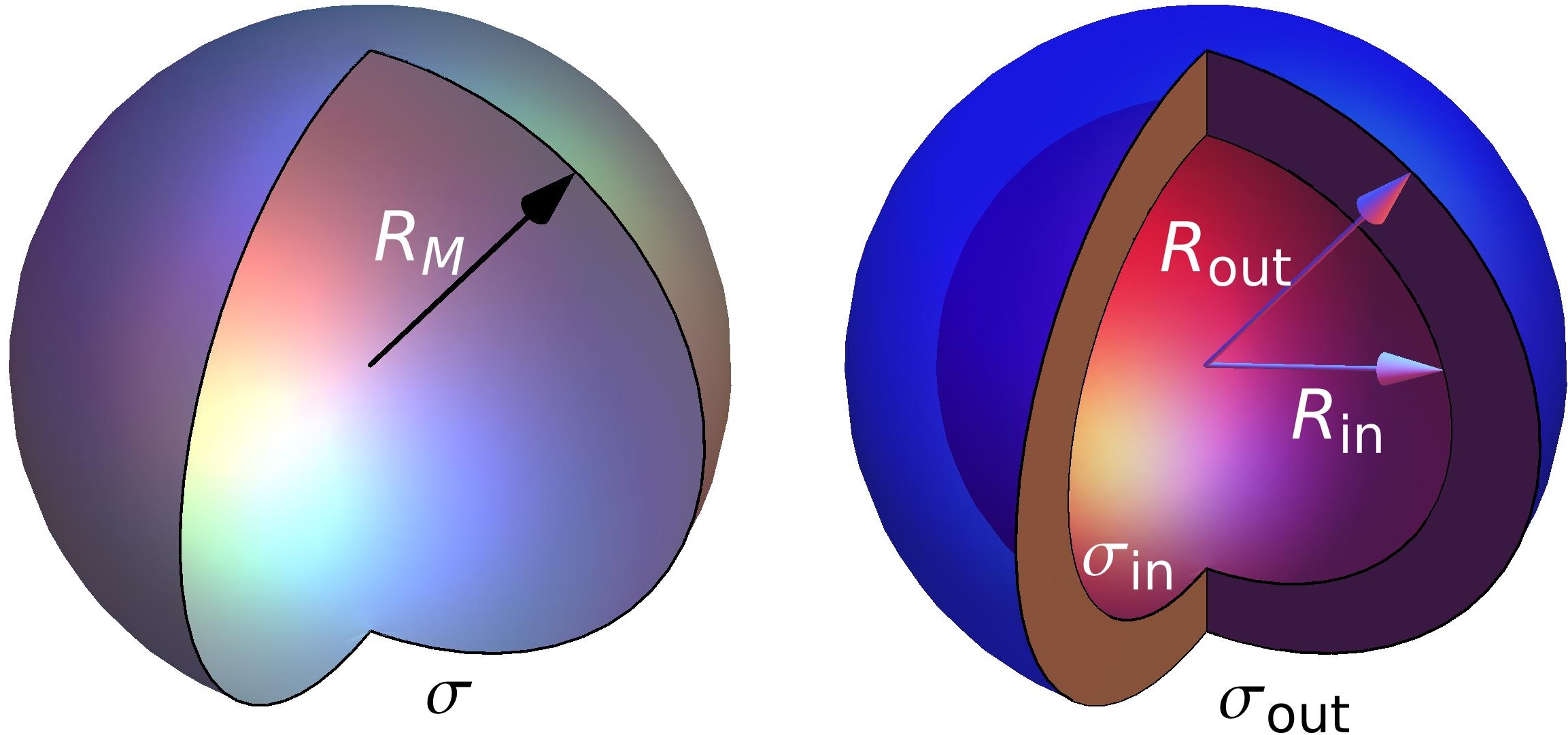}
\caption{Schematic representation of the single-shell and double-shell models treated in the paper. Left: single-shell model with mean radius $R_M$ and surface charge distribution $\sigma$. Right: double-shell model with the inner shell of radius $R_{in}$ and outer shell of radius $R_{out}$. The surface charge distributions pertaining to the two shells are denoted by $\sigma_{in}$ and $\sigma_{out}$.}
\label{fig:skica1}
\end{center}
\end{figure}

We focus on two simple models most widely used: a single, infinitely thin charged shell of radius $R_M$ and surface charge density $\sigma$ as shown in Fig.~\ref{fig:skica1}a \cite{Siber2007,Kegel2004}, and two thin shells of inner and outer radius $R_{in}$ and $R_{out}$ (giving a capsid thickness of $\delta_M=R_{out}-R_{in}$), carrying surface charges of $\sigma_{in}$ and $\sigma_{out}$ (Fig.~\ref{fig:skica1}b) \cite{Siber2007,Prinsen2010}. We will refer to the two models as the {\em single-shell} and {\em double-shell} model, respectively. Besides the monopole (total) charge distribution, we also consider the dipole distribution on such model capsids. The analysis is done solely for empty viral capsids not encapsidating any genetic material. 

In our analysis we use experimental data deposited in the VIPERdb database~\cite{VIPER}. This allows us to construct three-dimensional structures of viral capsids, from which we obtain the various mass and charge distributions within the capsid. We consider not only the distribution of atoms inside a capsid, but the distribution of amino acids (their positions taken as centers-of-mass of their constituent atoms) and complete protein chains as well.

Some capsid data do not contain the positions of all atoms but only the positions of alpha carbons -- in such cases we equate their positions with the positions of the amino acids to which they belong. Due to the methods of detection there are also no hydrogen atoms included in the experimental data. We have tested the effect of the lack of hydrogen atoms on our analysis by adding the hydrogen atoms via the MolProbity web server~\cite{MolProbity} to several different capsid entries. As expected, their effect on the mass distributions can be neglected, and we did so throughout our analysis.

To obtain the charge distributions of the capsids we extract the positions of charged amino acids from the experimental data by using Tcl scripting language in VMD~\cite{VMD}. At physiological \textsl{p}H of 7.4 we consider the following amino acids as charged~\cite{Betts2003}: aspartic acid (ASP) and glutamic acid (GLU) carrying a charge of $-1.0\;e_0$, lysine (LYS) and arginine (ARG) carrying a charge of $+1.0\;e_0$, and histidine (HIS) carrying a fractional charge of $+0.1\;e_0$ (where $e_0$ is the elementary charge).

The available experimental data cannot capture the usually disordered N-tails of proteins, which in certain cases do carry a significant charge~\cite{Belyi2006}. To estimate to what extent this affects our analysis we also compare the capsid protein sequences of viruses deposited in VIPERdb with the full sequences obtained from the UniProt database of protein sequences~\cite{UNIPROT}. 

In the following sections we extract and analyze the parameters of these simple models from the experimental data which look like the examples shown in Figs.~\ref{fig:example} and~\ref{fig:example3}. We analyze approximately 130 viruses from different families and compare their corresponding model charge and mass distribution parameters.

We classify the different viruses by their genome (single-stranded (ss) DNA and ssRNA on one hand and double-stranded (ds) DNA and dsRNA on the other)~\cite{Roos2010} and Caspar-Klug triangulation number $T$~\cite{Baker1999,Mannige2010}. These are the most conspicuous properties that classify the analyzed viruses; there are others, for example the secondary/tertiary structure of capsid proteins (i.e. presence of $\alpha$-helices, $\beta$-barrels, \ldots). However, we expect such additional properties play a smaller role in the task at hand~\cite{Speir2008}, and their inclusion would yield no additional insight in our analysis. We consider separately the bacteriophages (which come with either DNA or RNA genome), as well as the $T=p3$ capsids\footnote{The $T=p\mathrm{seudo}\,3$ icosahedral capsids do not obey the Caspar-Klug principle of quasi-equivalence because the basic unit is composed of three different (but morphologically similar) proteins.} of RNA viruses (which are abundant in our sample), since they might differ in their properties~\cite{Speir2008}.

\section{Single- and Double-shell Models}

We begin our analysis by constructing single-shell and double-shell models from the mass distributions in different viral capsids. The single, infinitely thin shell model is characterized by one parameter only, the mean capsid radius $R_M$ (Fig.~\ref{fig:skica1}a). The latter is extracted from the radial mass distribution in the capsid
\begin{equation}
\label{eq:masd}
\rho(r)=\frac{\Delta m}{4\pi r^2\Delta r},
\end{equation}
where the angular coordinates have already been projected out. This can be done for either the distribution of capsid atoms, centers-of-mass of amino acids, or centers-of-mass of proteins. The differences between these are within a couple of angstroms for most capsids, so we concern ourselves mainly with the distribution of capsid protein atoms.

The double-shell model on the other hand is characterized by two radii, the inner (hypotopal) and the outer (epitopal) radius $R_{in}$ and $R_{out}$ (Fig.~\ref{fig:skica1}b). Their difference is the capsid thickness $\delta_M = R_{out}-R_{in}$. These parameters are again obtained from the radial density distribution, with the thickness defined as the full-width-half-maximum (FWHM) of the distribution, and the inner and outer radius defined as the inner and outer half-maximum of the distribution. The bin size of the distribution influences the result to some extent, but the effect is still lower than the usual experimental precision. Also, since the exact half-maxima are never achieved due to the discreteness of the distribution, the condition they have to satisfy is to lie within 5\% around the half-maximum.

\begin{figure}[!tp]
\begin{center}
\includegraphics[width=\columnwidth]{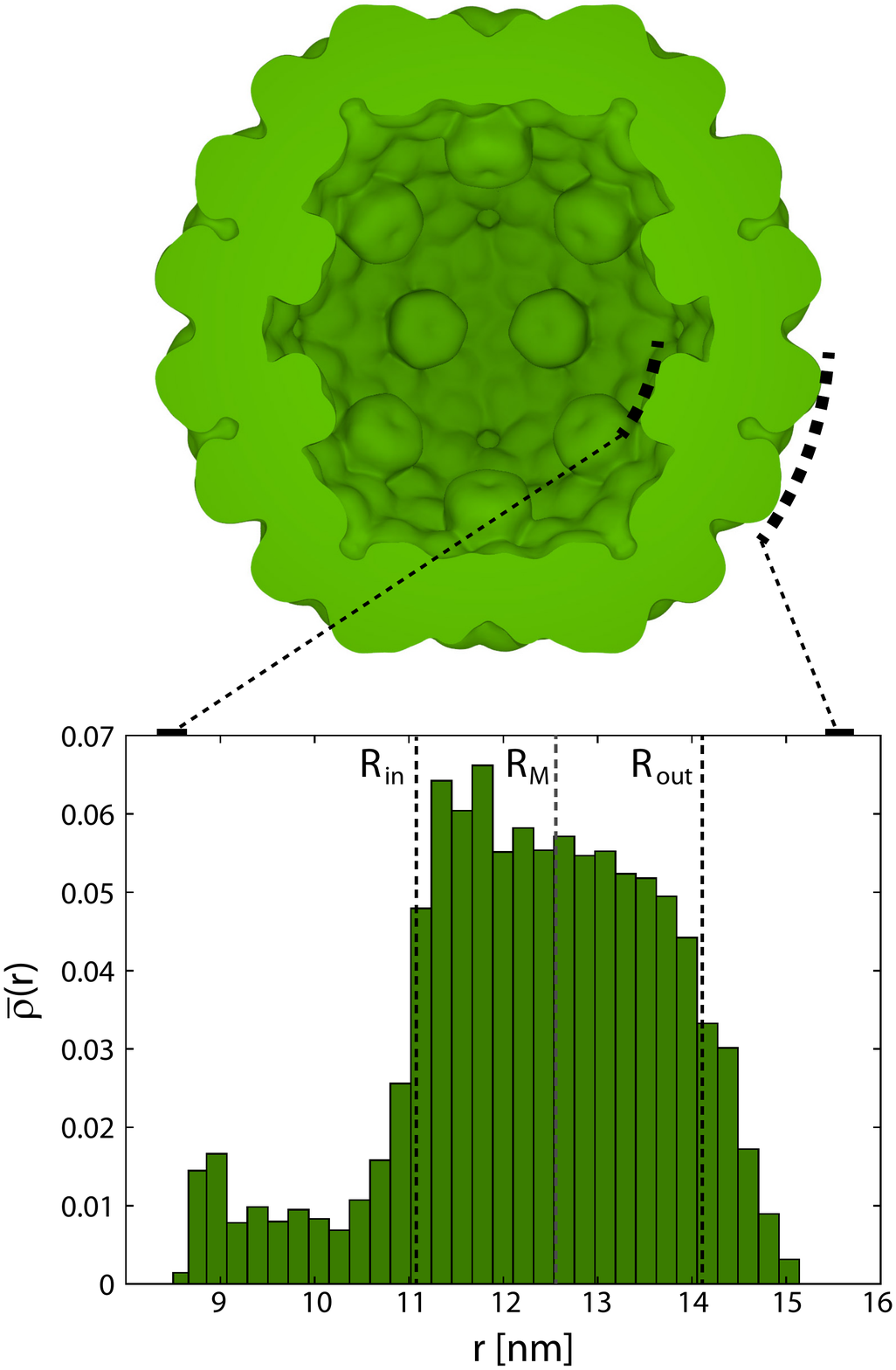}
\caption{Cross-section of (experimentally determined) capsid mass distribution in the example of the cucumber mosaic virus (ssRNA) capsid (strain FNY), constructed from RCSB Protein Databank entry 1f15. The drawing was constructed with a procedure described in Ref.~\cite{Siber2012} with $W = 1.34$ nm and $t = 0.85$, where all amino acids were assigned strength (``$q/e_0$'') 1. Protrusions can be seen on the capsid interior which are the roots of protein N-tails; comparison with the full protein sequence shows that they are not complete. The inset shows the radial mass distribution (Eq.~\ref{eq:masd}) across the capsid, normalized so that total area of the histogram equals 1; marked are the mean capsid radius $R_M$ (single-shell model) and the inner and outer radii $R_{in}$ and $R_{out}$ (double-shell model).}
\label{fig:example}
\end{center}
\end{figure}

To illustrate how this analysis is done we consider the example of cucumber mosaic virus (CMV, PDB ID 1f15). Figure~\ref{fig:example} shows the radial mass distribution in the capsid, where we can see that the root parts of the protein N-tails, prominent in this example, are protruding into the capsid interior as defined by the hypotopal radius of the distribution $R_{in}$. Any significant outer protrusions such as spikes are located in the exterior of the capsid as defined by the epitopal radius $R_{out}$. These details are not included in the simpler single-shell model, characterized only by the mean capsid radius $R_M$.

\begin{figure}[!htp]
\begin{center}
\includegraphics[width=\columnwidth]{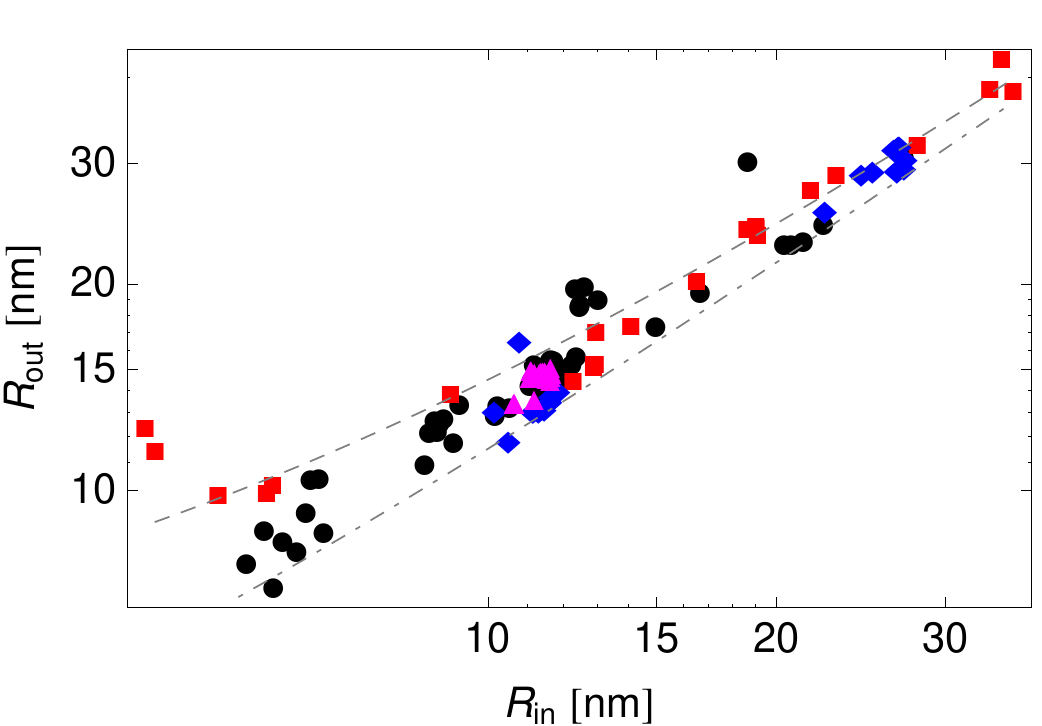}
\caption{Outer capsid radius compared to inner capsid radius of the double-shell model. The thickness of the capsid emerges naturally from this linear dependence: the dashed line shows a thickness of 4.5 nm (i.e. $R_{out}=R_{in}+4.5\ \mathrm{nm}$), and the dot-dashed line shows a thickness of 1.5 nm. Approximately two-thirds of the analyzed capsids have a thickness between 2-4 nm. Symbols encode some different virus types: single-stranded genome (circles), double-stranded genome (squares), bacteriophages (diamonds), and $T=p3$ ssRNA viruses (triangles). Same symbols are used throughout the paper in other similar figures.}
\label{fig:rinrout}
\end{center}
\end{figure}

In Fig.~\ref{fig:rinrout} we next plot the inner and outer radius of the double-shell model for the entire dataset of analyzed viruses\footnote{In all such plots the following legend is used for different virus types: single-stranded genome (circles), double-stranded genome (squares), bacteriophages (diamonds; both ss and ds genome), and $T=p3$ ssRNA viruses (triangles).}. Capsid thickness naturally follows from the apparent linearity of their relation and is generally well defined. For more than $75\%$ of viruses in our sample the thickness is confined to a narow range, $\delta_M\sim1.5$-$4.5\ \mathrm{nm}$.

To a good approximation, the mean capsid radius of the single-shell model increases with the square root of the capsid $T$-number, which means that one can idealize the capsid as consisting of uniformly distributed copies of a disk-shaped (or prism-shaped) elementary protein with a fixed area. A minimal model of this type for equilibrium capsid structure with explicit interaction between capsomeres on a spherical shell has received much attention recently~\cite{Marzec1993,Zandi2004}.

\begin{figure}[!htp]
\begin{center}
\includegraphics[width=\columnwidth]{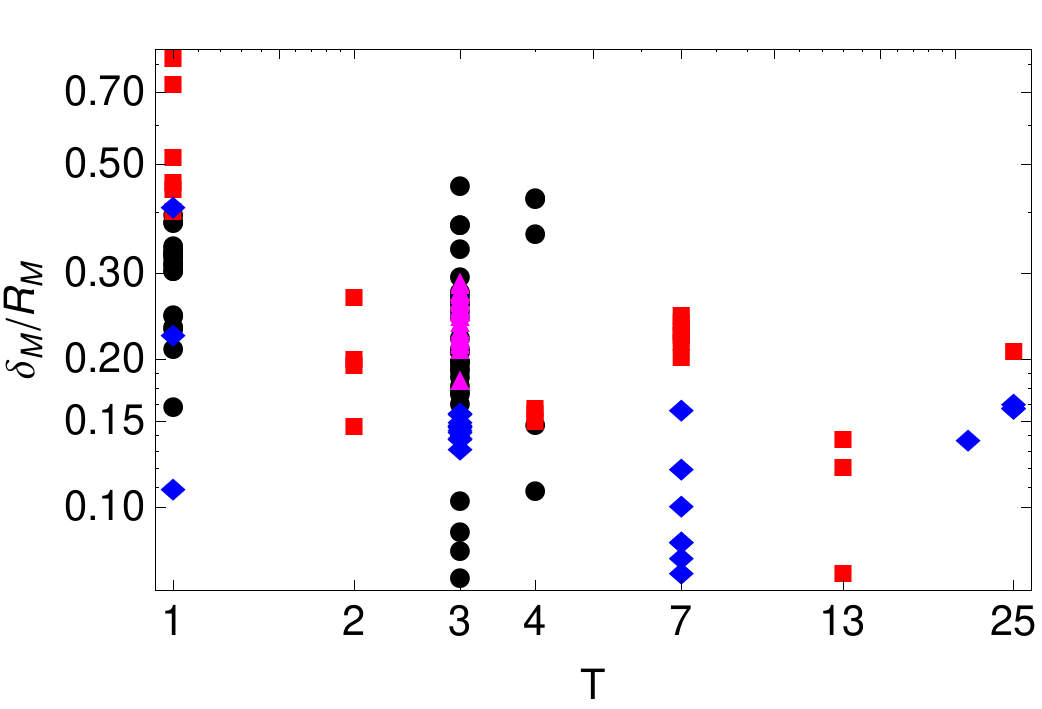}
\caption{Ratio of capsid thickness and capsid mean radius $\delta_M/R_M$ as a function of the triangulation number. The capsid thickness becomes more and more comparable to the capsid radius as the triangulation number gets lower.}
\label{fig:fwhmT}
\end{center}
\end{figure}

An additional point of interest is also the ratio of the capsid thickness and the mean capsid radius $\delta_M/R_M$, as this can influence the validity of mechanical models of viruses, for instance continuum elasticity models of thin elastic shells~\cite{Roos2010,Siber2009}. Analysis of this ratio is shown in Fig.~\ref{fig:fwhmT}. For the average virus analyzed this ratio lies around $0.2$, but is (expectedly) no longer small for smaller, $T=1$ viruses, where the idealization of a thin protein shell is misleading.

These characteristics of the capsid architecture turn out to be insensitive to taking the mass distribution instead of the position distribution, which barely affects the calculated mean radius or the thickness of the capsid. A more detailed analysis of the conserved geometrical properties of viruses and their elastic properties will be published elsewhere (A. Lo\v sdorfer Bo\v zi\v c, A. \v Siber, and R. Podgornik, in preparation).

\section{Charge Distributions}

As specified in Section~\ref{sec:s2m}, there are five amino acids in proteins that carry charge at physiological \textsl{p}H. However, there is some uncertainty as to whether these ionizable amino acids are charged or not when buried inside a protein. Either the dissociation cost  for charges buried in the protein interior is too high and the buried charges are therefore virtually absent~\cite{Ptitsyn2002}, or the converse is true and the majority of ionizable amino acids buried inside the protein are ionized~\cite{Gunner2006}. Yet another possibility is that the local environment of a buried ionizable amino acid is changed, so that its charge is modified~\cite{Isom2008,Isom2010}.

With this in mind we consider two limiting cases: in the first one, we take all the ionizable amino acids as charged, no matter where they are located. In the second case we consider as charged only the ionizable amino acids lying on the periphery of the capsids as defined by their inner and outer radii. This is admittedly a simplified picture, but it enables us to cover the extreme cases. Only a complete \emph{ab initio} quantum chemical calculation of the electronic properties of capsid proteins in contact with aqueous solvent and neighboring proteins could resolve the issue of the correct charging model for the amino acids \cite{WYChing2011, Pichierri}.

\begin{figure}[!htp]
\begin{center}
\includegraphics[width=\columnwidth]{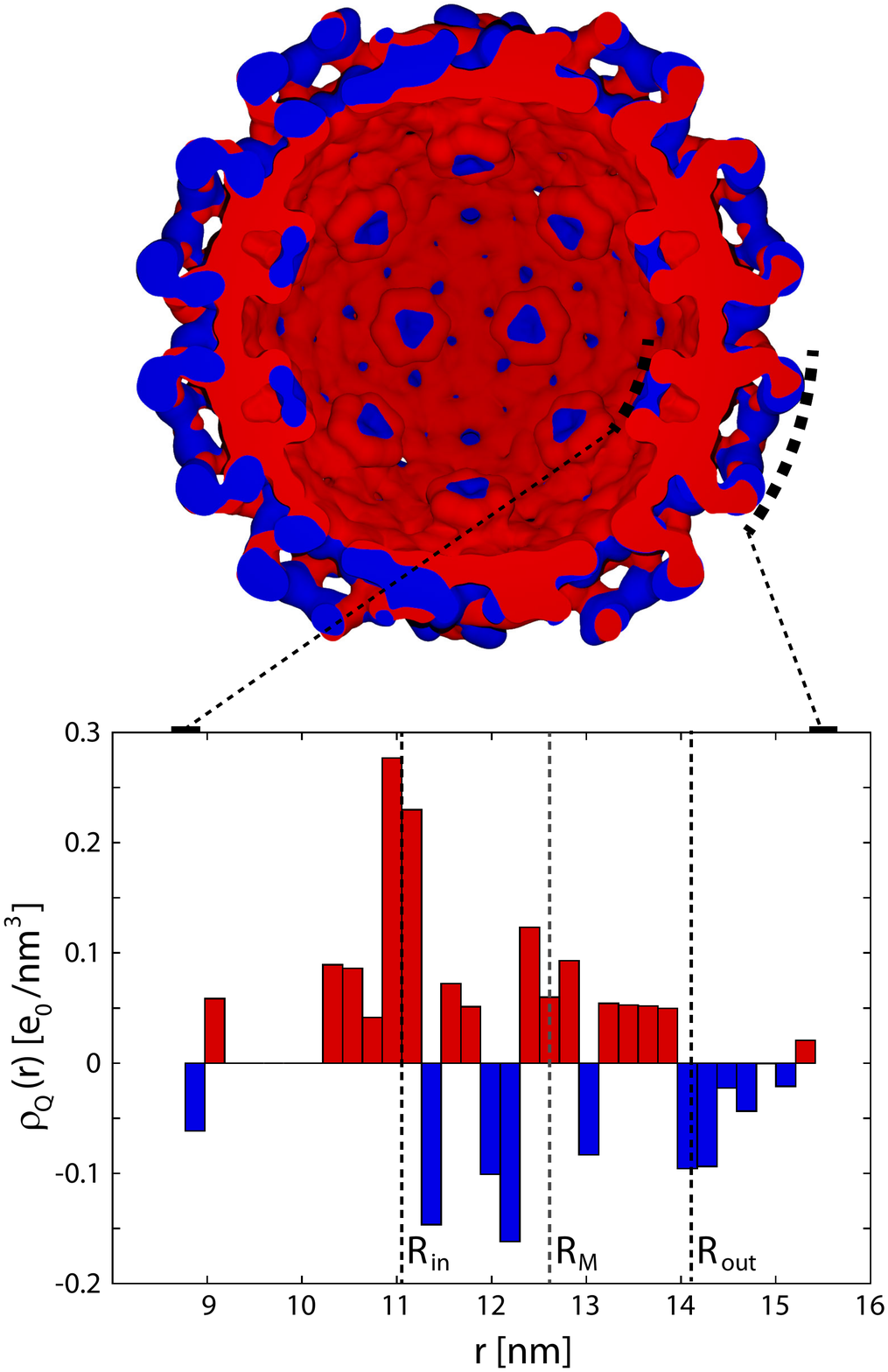}
\caption{Cross-section of charge distribution in the example of cucumber mosaic virus (PDB ID 1f15). The 3D representation is constructed as described in Ref.~\cite{Siber2012} with $W = 1.34$ nm and $t = 0.85$. The histogram plot shows corresponding radial charge distribution across the capsid. Note that the 3D representation separately represents negative (blue) and positive (red) charge densities, while the histogram shows the total charge density distribution, calculated by weighing both charge distributions. As the negative and positive charge distributions overlap, in order to clearly show both of them, the positive and negative distributions are infinitesimally shifted with respect to each other, so that on the right (left) half of the 3D representation the positive (negative) distribution is infinitesimally closer to the viewer. Marked are the capsid mean mass radius $R_M$ and the inner and outer radii $R_{in}$ and $R_{out}$ of the single- and double-shell models, respectively.}
\label{fig:example3}
\end{center}
\end{figure}

\begin{figure}[!htp]
\begin{center}
\includegraphics[width=\columnwidth]{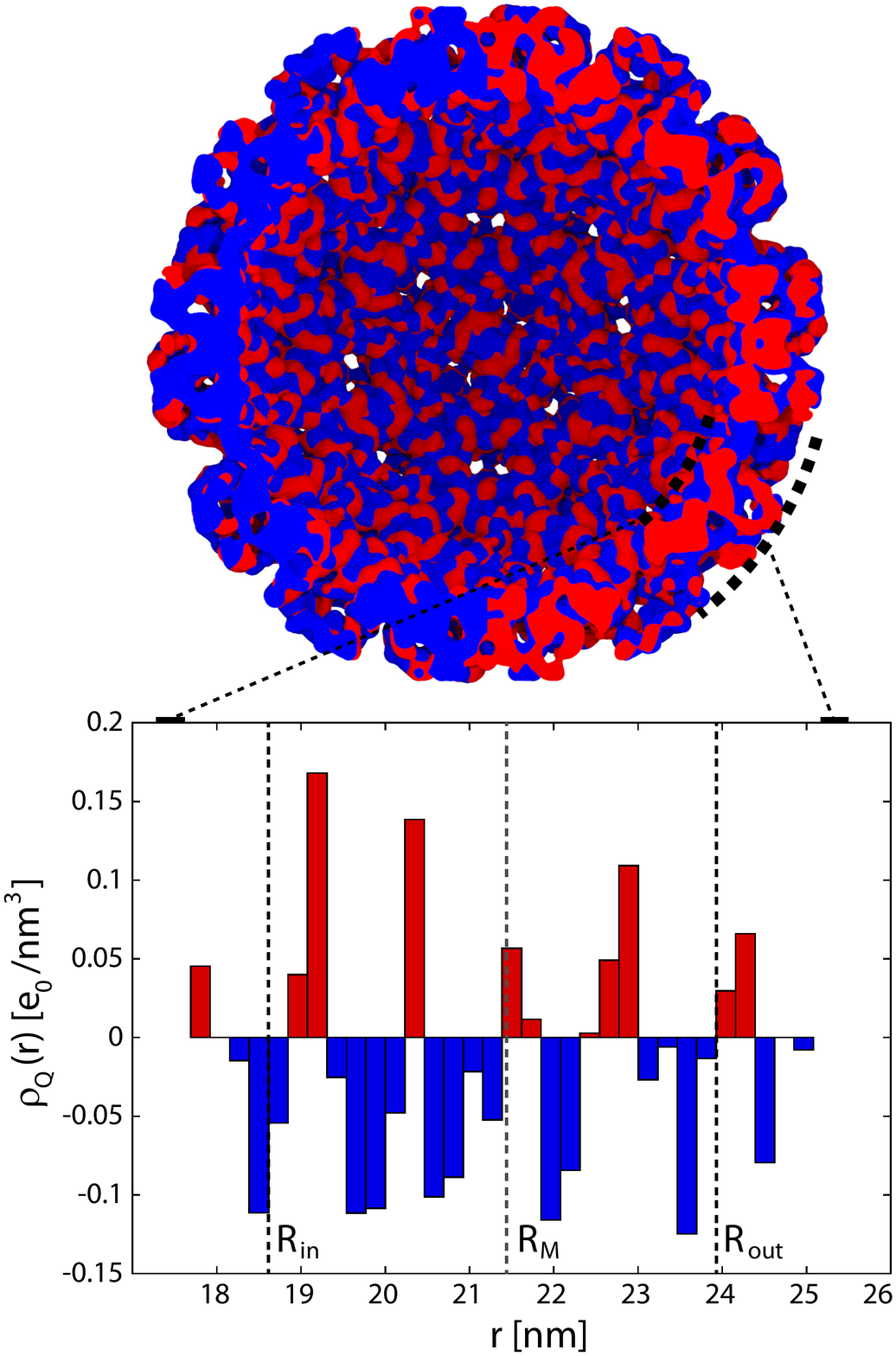}
\caption{Cross-section of charge distribution in the example of simian virus 40 (PDB ID 1sva). The figure is constructed in the same manner as Fig.~\ref{fig:example3}. However, the radial charge distribution in this example cannot be easily categorized, and, most notably, does not have a pronounced positively charged inner part of the capsid and negatively charged outer part of the capsid. Marked are again the capsid mean mass radius $R_M$ and the inner and outer radii $R_{in}$ and $R_{out}$ of the single- and double-shell models, respectively.}
\label{fig:example4}
\end{center}
\end{figure}

A sample radial charge distribution is again shown for the CMV in Fig.~\ref{fig:example3}. All the ionizable amino acids are taken as charged, regardless of their position in the capsid. In this case, we observe that the charges on the hypotopal and epitopal surfaces are mostly positive and mostly negative, respectively; there are also some charges buried in the capsid wall. These are the only distinguishing features of an otherwise very complicated charge distribution. The distribution of charges in the capsid can vary significantly from virus to virus, and there appears to be no simple way of classifying them. One example of a very different distribution is shown in Fig.~\ref{fig:example4} for the case of simian virus 40 (PDB ID 1sva). Here, it is difficult to separate the charge distribution into a positively charged hypotopal surface and a negatively charged epitopal surface, and there is a good deal of charge variation within the capsid wall.

\subsection{Total Charge}

The total charge of the capsid $Q$ is calculated as the sum of all charged amino acids in the capsid, $Q = \sum_i q_i$, within the two limiting models described above. We also introduce the mean radius of the distribution of absolute charge (mean charge radius) $R_Q$,
\begin{equation}
\label{eq:rq}
R_Q=\frac{\sum_i |q_i| r_i}{|Q|},
\end{equation}
where $q_i$ are the charges of amino acids located at radii $r_i$. The charge mean radius of most of the viruses differs from the mass mean radius by up to a few percent.

The total charge of the single-shell model is usually given in terms of a surface charge density $\sigma$:
\begin{equation}
\label{eq:chr}
\sigma=\frac{Q}{4\pi R_M^2};
\end{equation}
this is the surface charge density of the single-shell model with all the ionizable amino acids being charged. Here we could equally well use $R_Q$ (Eq.~\ref{eq:rq}) instead of $R_M$ but we stick with the latter for consistency.

The surface charge densities of the double-shell model $\sigma_{in}$ and $\sigma_{out}$ are similarly defined as
\begin{equation}
\label{eq:chr2}
\sigma_{in/out}=\frac{Q_{in/out}}{4\pi R_{in/out}^2}.
\end{equation}
The charge on the inner shell is $Q_{in}=\sum_i q_i\;;\;r_i(q_i)<R_{in}$, and the charge on the outer shell is calculated in an analogous fashion. In order to compare the total charge of the two models, we also define
\begin{equation}
\label{eq:chIO}
\sigma_{IO}=\frac{Q_{in}+Q_{out}}{4\pi R_M^2}.
\end{equation}
This can be considered as the surface charge density of the single-shell model with only peripheral amino acids (i.e. not buried inside the capsid as defined by the double-shell model) taken as charged.

\begin{figure}[!htp]
\begin{center}
\includegraphics[width=\columnwidth]{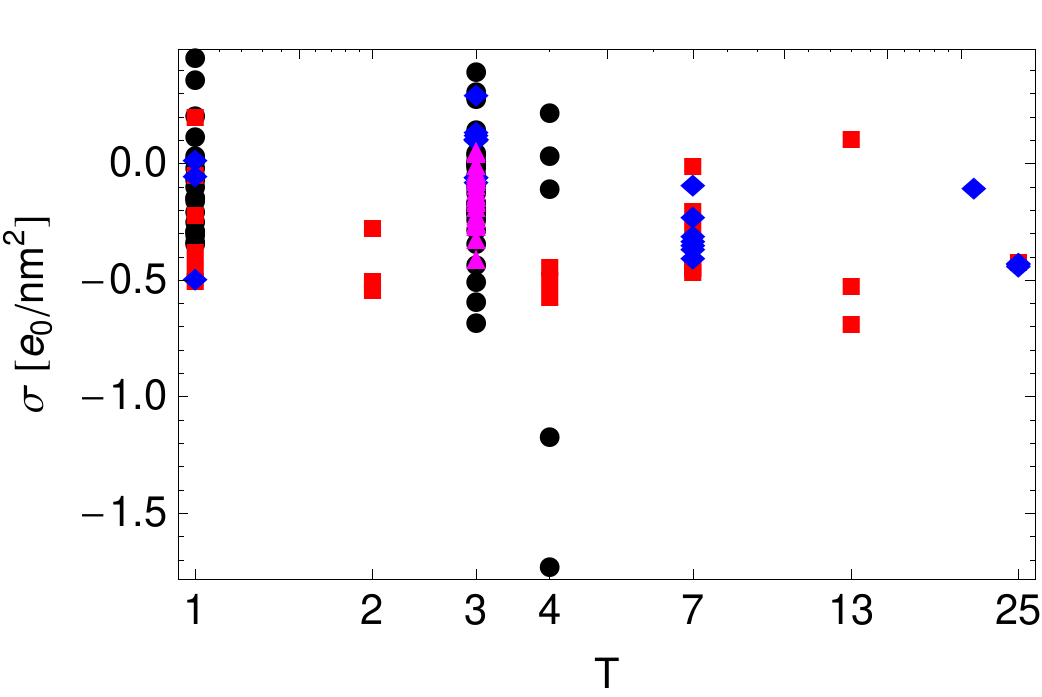}
\includegraphics[width=\columnwidth]{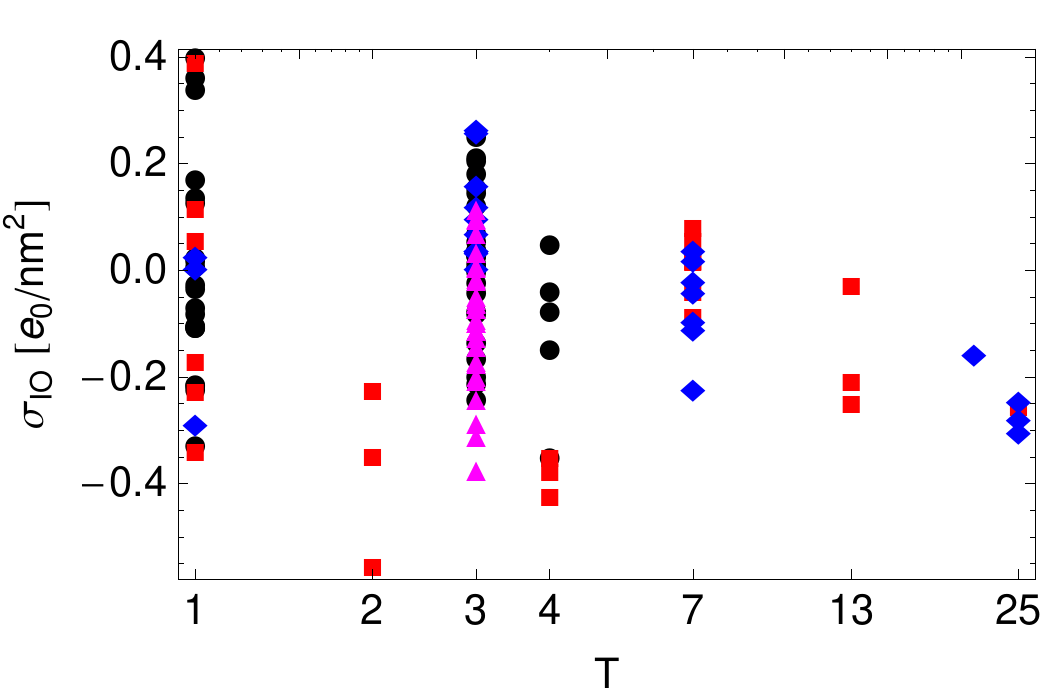}
\caption{Distributions of the total capsid surface charge density depending on the triangulation number, taking into account either all the charged amino acids ($\sigma$; Eq.~\ref{eq:chr}) or only the charged amino acids lying outside the mean capsid thickness ($\sigma_{IO}$; Eq.~\ref{eq:chIO}). In the latter case the capsids tend to carry slightly more positive charge; the relevant range of surface charge densities is in both cases well described by the interval $[-0.4,0.4]\ e_0/\mathrm{nm}^2$.}
\label{fig:sigma}
\end{center}
\end{figure}

The dependence of the total charges of the single-shell model in both limits ($\sigma$ and $\sigma_{IO}$) on the capsid $T$-number is shown in Fig.~\ref{fig:sigma}. We can see that for the majority of viruses the total charge becomes more positive when we exclude the buried charges. The values of capsid surface charge densities mostly lie within the range from $-0.4$ to $+0.4$ $e_0/$nm$^2$. Invoking the previously obtained $R_M$ this implies net charge values in the range $|Q| \lesssim 4500$ $e_0$. Empty viral capsids are obviously quite charged and their interactions either between themselves or with other structural components of the cell must be to a large extent modulated by electrostatics.

\begin{figure}[!bhp]
\begin{center}
\includegraphics[width=\columnwidth]{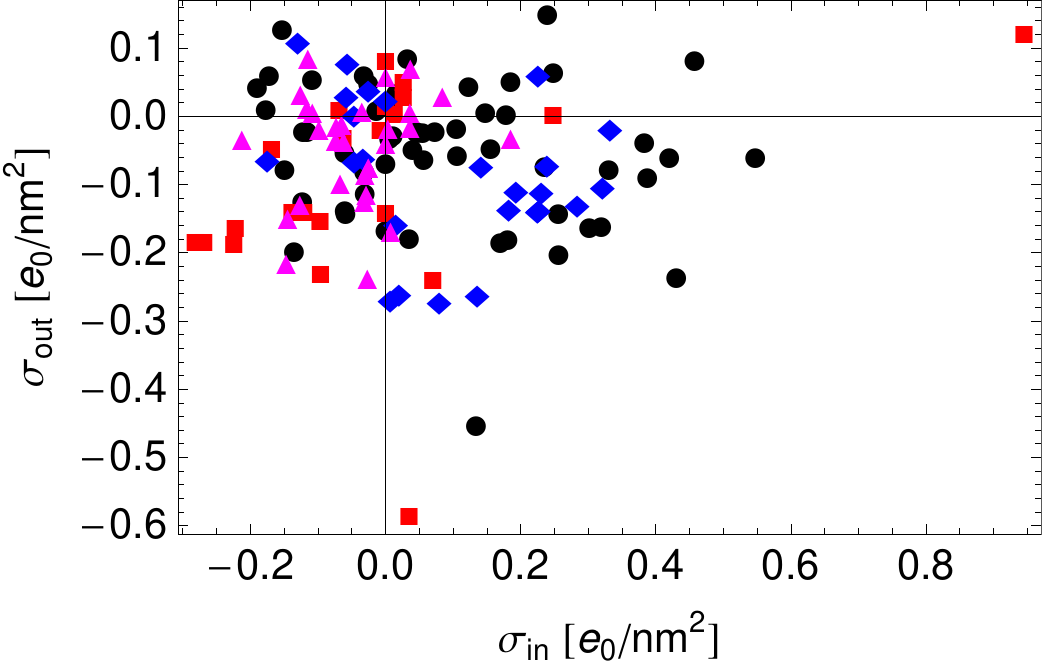}
\includegraphics[width=\columnwidth]{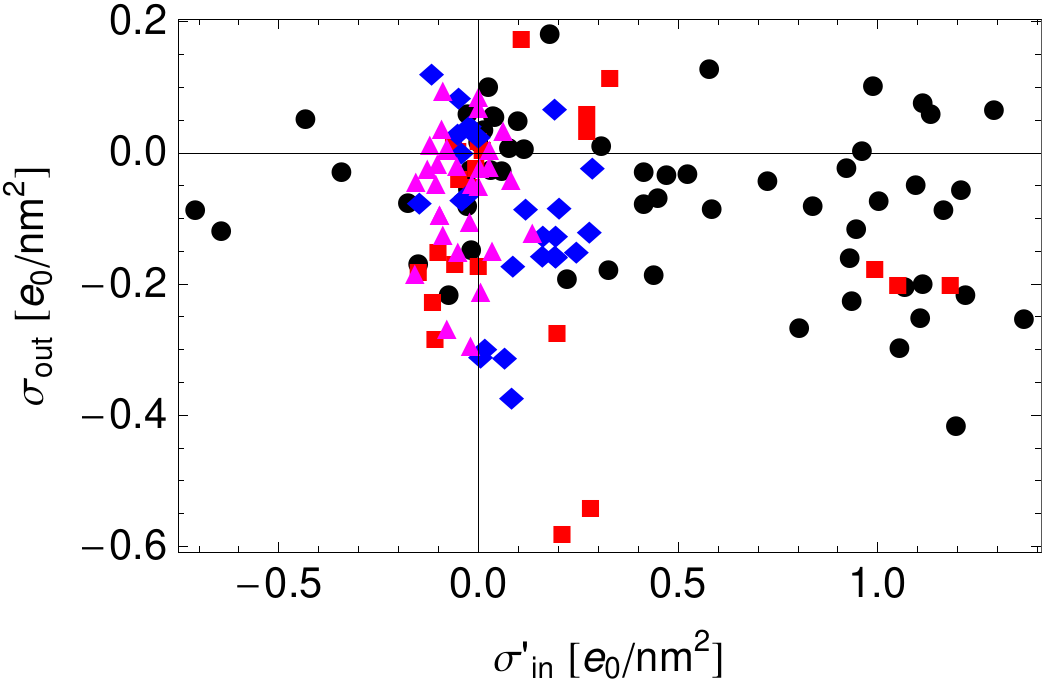}
\caption{Top panel: comparison of the surface charge densities on the inner and outer shells of capsids (Eq.~\ref{eq:chr2}). The majority of the viruses tend to have at least slightly negatively charged outer shell. There is more diversity concerning the charge on the inner shell, which is in our sample centered around zero net charge, with viruses having either negatively or positively charged interior. Bottom panel: same as above, with added disordered N-tails of the proteins. There is a noticeable shift of the inner shell charge (to which the missing sequences were attributed) towards more positive values.}
\label{fig:sigIO}
\end{center}
\end{figure}

\begin{figure}[!htp]
\begin{center}
\includegraphics[width=\columnwidth]{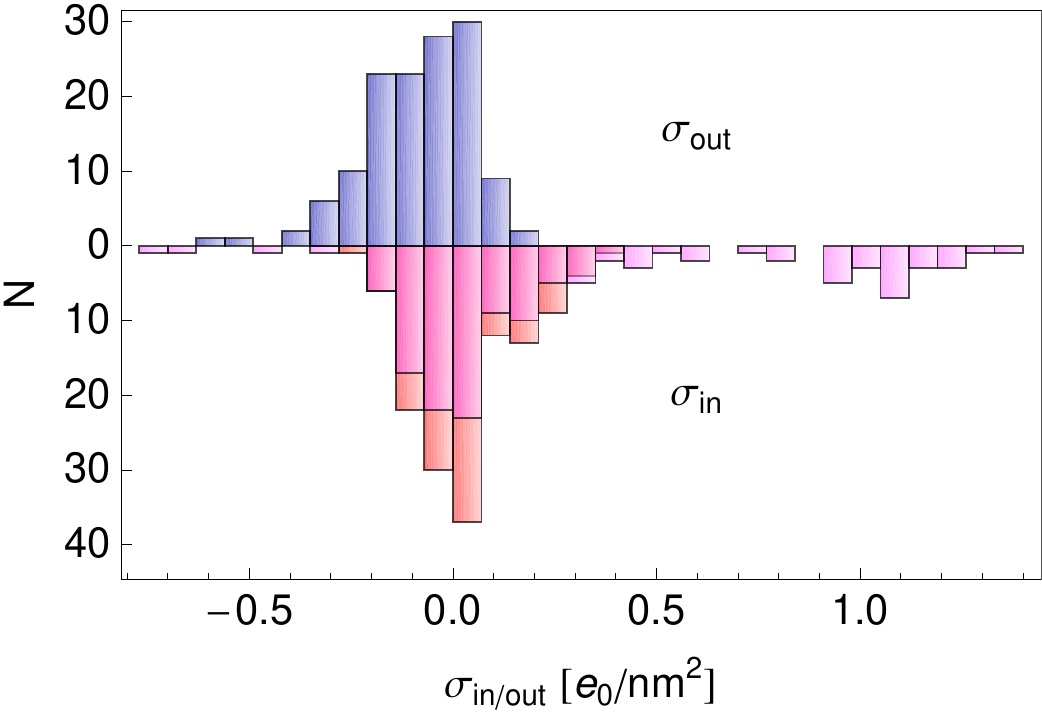}
\caption{A histogram showing the distribution of inner and outer surface charge densities of the double-shell model in the sample of viruses used in the analysis. The upper part shows the outer surface charge density (blue), and the lower part shows the inner surface charge density without (red) and with added charge of the N-tails (magenta).}
\label{fig:histogram}
\end{center}
\end{figure}

In Fig.~\ref{fig:sigIO} we then compare the inner and outer surface charge densities of the double-shell model. An emerging feature, which can be also discerned from the histogram in Fig.~\ref{fig:histogram}, is that the outer charges of viruses are close to zero or slightly negative; on the contrary, there are quite some viruses that carry a significant positive inner charge, even though a lot of them still carry an inner charge close to zero. The viruses carrying a positive inner charge in this case are mostly viruses with single-stranded genome (with the exceptions of $T=1$ and $T=p3$ capsids) as well as bacteriophages with single-stranded genome.

\subsection{Effect of Missing (Disordered) N-tails}

The basic (positively charged) N-tails of capsid proteins are largely unresolved in X-ray scattering experiments and Belyi and Muthukumar~\cite{Belyi2006} have shown that due to their positive charges they can strongly interact with the oppositely charged RNA genome. This interaction is also a major factor in constraining the length of viral genome, implying a linear relation between the number of positive charges on the tails and the length of the encapsidated RNA \cite{Belyi2006,Hu2008,Siber2008}.

The effect of missing disordered tails in the experimental structure data can be most easily estimated from the changes in the total capsid charge brought about by the positively charged N-tails. The missing charge is calculated from the full primary sequences of capsid proteins. Since nothing can be said about their position (other than that they are most likely disordered and located on the hypotopal side of the capsid), we take all the missing charges to be located in the interior of the capsid, that is on the inside of $R_M$ or $R_{in}$ within the single- and double-shell models, respectively. This is an assumption which should hold true for most of the analyzed viruses, but cannot be easily verified.

By adding the charge contributed by the N-tails we get an estimate of the charge correction $\Delta Q$ and from there the new values for the total surface charge density $\sigma'$ in the single-shell model and new values for the inner surface charge density $\sigma_{in}'$ in the double-shell model. From the latter we can also obtain the corrected total surface charge density of Eq.~\ref{eq:chIO}, $\sigma_{IO}'$; all the surface charge densities are again normalized with $R_M$.

\begin{figure}[!hbp]
\begin{center}
\includegraphics[width=\columnwidth]{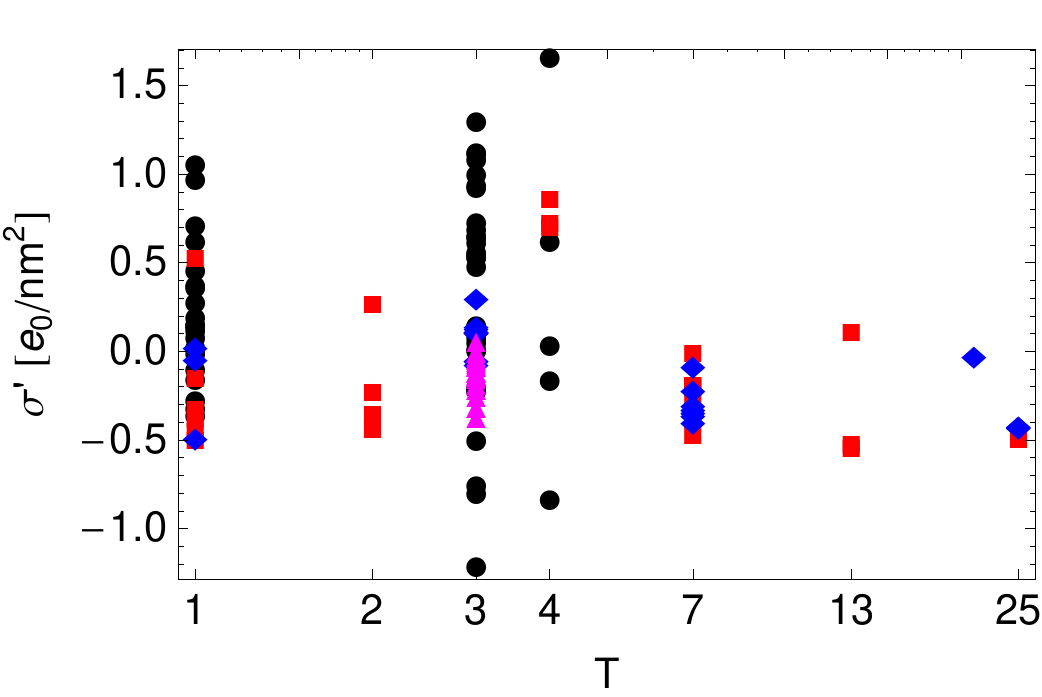}
\includegraphics[width=\columnwidth]{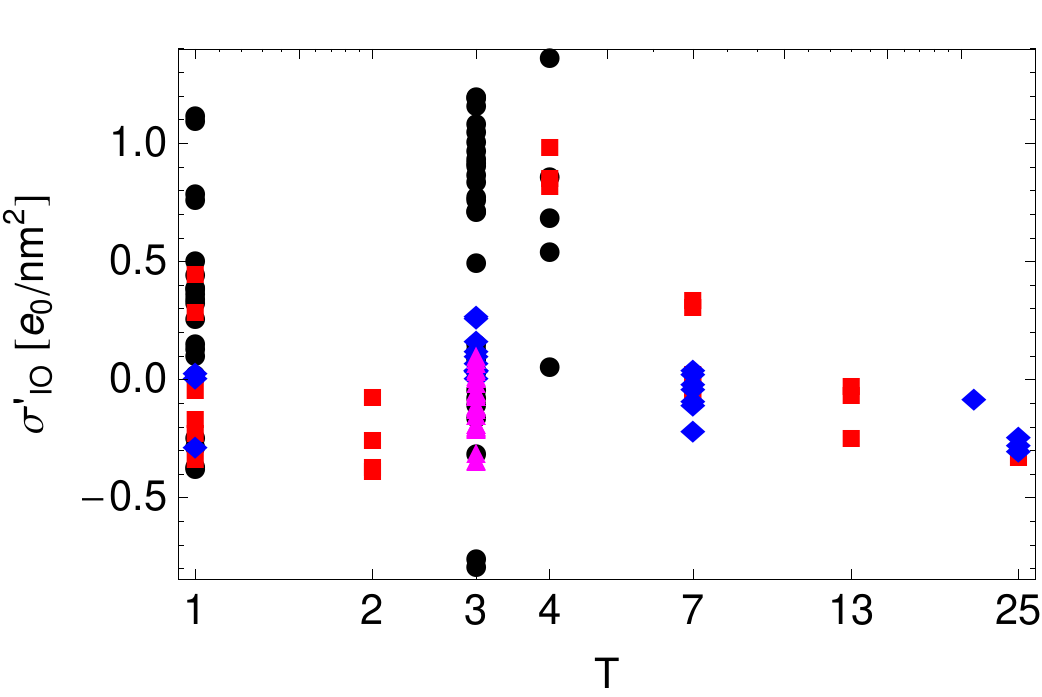}
\caption{Surface charge density of the capsids with added disordered N-tails of the proteins plotted against the triangulation number, for both limiting cases considered (compare with Fig.~\ref{fig:sigIO}). The total charge moves towards more positive values in both cases; this trend is least pronounced in bacteriophages and $T=p3$ ssRNA viruses.}
\label{fig:sigmod3}
\end{center}
\end{figure}

The distributions of the new surface charge densities of the single-shell model as a function of the triangulation number are shown in Fig.~\ref{fig:sigmod3}. In general, a trend toward more positive charge is observed by corrections up to $|\Delta Q|\sim6000\ e_0$. The same is true also for the double-shell model where the total surface charge density decomposes into the hypotopal in epitopal contributions (Figs.~\ref{fig:sigIO} and~\ref{fig:histogram}). The rationalization for this rescalings of the capsid charge once one adds explicit charges on the disordered N-tails could be envisioned as stemming primarily from the strong N-tail genome electrostatic interactions \cite{Belyi2006}. Since the genome is negatively charged, the hypotopal N-tails effectively act locally to completely screen this charge, conferring much needed stability to the virus.

The most pronounced and consistent changes can be observed in the case of viruses with single-stranded genome, with a clear separation of the total charge between $T=3$ single-stranded viruses and the rest, and a slightly less pronounced separation in the $T=1$ viruses as well. The charges of the bacteriophages remain mostly unchanged after the explicit addition of N-tail charges, as do the charges of $T=p3$ ssRNA viruses. The latter case is somewhat surprising, as the majority of single-stranded viruses undergo an increase of charge. The effect on double-stranded viruses is not so systematic.  

From these results we conclude that the surface charge of the capsid is quite large, being comparable to the equivalent surface charge of a DNA molecule. In absolute terms the number of effective charges can go into tens of thousands, which is an impressive charge even after all the screening and condensation effects are taken into  account, making viral capsids quintessential charged nano-objects~\cite{Siber2012}. The electrostatic interactions stemming from this huge capsid charge are therefore important and cannot be neglected.

\subsection{Dipole Distribution}

Lastly, we analyze the first higher order multipolar moment of the capsid charge distribution, the dipole moment. The electric dipole of the capsid shell is defined as
\begin{equation}
\mathbf{P}(\mathbf{r}_0)= \sum_i q_i(\mathbf{r}_i-\mathbf{r}_0).
\end{equation}
where $q_i$ are again the charges of amino acids located at radii $\mathbf{r}_i$ within the capsid shell.
The dipole distribution is not invariant with respect to geometric description and has to be calculated with respect to some particular reference point $\mathbf{r}_0$~\cite{Jackson1999}. We choose for the origin the radius of the centre of absolute charge $\mathbf{R}_Q$. Apart from the absolute magnitude of the dipolar moment we again consider the surface dipole density, normalized with the capsid mean radius. The surface dipole density is completely analogous to the surface charge density introduced before. Since the dipolar moment and its local surface density are vectors, we can decompose them into a radial and a tangential component -- across and along the capsid wall -- and compare their respective magnitudes.

We calculate the dipolar moment for the basic asymmetric unit of a capsid: the conglomeration of a $T$-number of proteins which, upon applying 60 rotation matrices of the icosahedral group, compose the entire capsid. This is done to simplify the analysis and enable us to make a good comparison of the results; in principle it would be possible to calculate the dipolar moment of each capsid protein, but we believe this would not serve any additional purpose in our analysis. It would even make sense to calculate the dipolar moments of either dimers, trimers, pentamers, hexamers, or whatever the basic structural units of each capsid is~\cite{Petsko2004}, so as to see if the dipolar moment plays a role in their interaction. However, these units differ from virus to virus, and would be difficult to address within our approach. In any case, we find that the magnitudes of the dipolar moments in capsid proteins are small, and these effects are thus likely to be small as well.

The majority of viruses have small surface dipole densities, below $0.02$ $e_0$/nm. For comparison, one could note that the surface dipole density of a completely oriented layer of water molecules at close packing would be $0.55$ $e_0$/nm. The obvious conclusion then is that if there is any ordered water on the periphery of the capsid, its effect will overwhelm the intrinsic dipolar moments of the capsid proteins. Note however that the surface water ordering in ``hydration layers'' would be highly contingent on the local protein charge distribution ~\cite{Hydration}. One should nevertheless remark here that the dipolar moment calculated above does not take into account the complete electronic structure of the proteins with implied partial charges within the protein cores that may eventually contribute to the total dipolar moments of the capsid proteins. Regardless, compared to monopolar, the dipolar surface charge density seems to be much less important.

\section{Discussion \& Conclusions}

We have performed a detailed statistical analysis of mass and charge distributions in approximately 130 empty viral capsids, and extracted the relevant parameters needed to construct simple single- and double-shell models of them. The complete list of analyzed viruses, their (triangulation) $T$-numbers, and genome types, as well as a compilation of the results presented in the paper, is available from the authors upon request.

The analysis of the charge distribution in capsids was based on several assumptions that do not have a universal validity, but are at present necessary to take as given. In structuring our models we ignored the dependence of the dissociation constants of amino acids on the detailed molecular environment as it would, even though possible in a case-by-case analysis~\cite{Langlet2008}, make our general approach completely untransparent. Nevertheless, these features should be investigated in the context of an improved model that would consider fully dissociated charge of amino acids on solvent accessible surface of a protein as well as the rearrangement of charges inside the protein due to quantum electron charge transfer~\cite{WYChing2011}. However, the calculation of the latter  is at present not feasible for such a large number of amino acids, and we thus focused only on the dissociated charge. Some viruses are also reported stable under \emph{in vitro} conditions at non-physiological {\sl p}H~\cite{Hirth1976}. Apart from the fact that this attests to the importance of electrostatic interactions in self-assembly of viruses it also has implications for their charges.

Therefore, some approximation for calculating the dissociation charges of amino acids in a protein has to be made, and it can be done in several ways~\cite{VIPER,Felder2007,Belyi2006}. We chose a straightforward and simple method for extracting the charges from 3D experimental data at a single value of solution {\sl p}H that enabled us to perform a consistent and general analysis. It is only one possibility though, and different approaches can yield quantitatively different results especially if the solution {\sl p}H variation is considered in full.

Within the limitations described above, we were able to quantify the radial capsid charge distribution, its corresponding  surface charge densities, dipole moments, and some of their geometric properties. This is clearly an important information to be had when using simple models of viral capsids. The monopolar surface charge density of the capsids was found to be quite large when compared with other charged biomolecules, being in the range $[-0.4,0.4]\ e_0/\mathrm{nm}^{2}$. We have also shown that for the overall charge of the virus capsids the disordered N-tails contribute significantly to the net charge, often changing its sign. Consequently, this also results in strongly positively charged interiors of ssRNA viruses, for which it has been suggested that the interior charge is correlated with the genome length~\cite{Belyi2006,Ting2011,Dragnea2012}.

While the dipolar charge contribution turned out to be on the other hand overall much smaller, it can nevertheless play an important role whenever stabilization of high energy structures hinges on important subdominant contributions. It is in fact this secondary dipolar density that most probably governs the short range interactions between capsomeres ~\cite{Hydration}.

Contrary to some of the capsid geometrical properties, the distribution of capsid charges does not seem to possess any regularity among viruses with similar triangulation numbers, genome types, or species, as was also observed by Michen and Graule~\cite{Michen2010} in the study of their isoelectric points. The choice of the dataset used in such a study can certainly influence the result to some extent~\cite{Ting2011}, and a future increase in the number and variety of available experimental data would undoubtedly improve the analysis.

\begin{acknowledgements}
One of us (A.L.B.) thanks A. Ljubeti\v{c} for introducing him to Tcl scripting language in VMD.

A.L.B. acknowledges the support from the Slovene Agency for Research and Development under the young researcher grant. A.\v{S}. acknowledges support from the Ministry of Science, Education, and Sports of Republic of Croatia (Grant No. 035-0352828-2837). R.P. acknowledges support from the Slovene Agency for Research and Development through research program P1-0055 and research project J1-4297.
\end{acknowledgements}

\bibliography{references_new}

\end{document}